\documentclass[preprint,amsmath,amssymb,aps,prb,lettersizepaper,superscriptaddress,longbibliography]{revtex4-1}
\usepackage{graphicx}

\begin{document}


\title{\large{Dipolariton formation in quantum dot molecules strongly coupled to optical resonators}}

\author{Marlon S. Dom\'inguez}
\affiliation{Grupo de F\'isica Te\'orica y Computacional, Escuela de F\'isica, Universidad Pedag\'ogica y Tecnol\'ogica de Colombia (UPTC), Tunja 150003, Boyac\'a, Colombia.}
\author{David F. Macias-Pinilla}
\affiliation{Grupo de F\'isica Te\'orica y Computacional, Escuela de F\'isica, Universidad Pedag\'ogica y Tecnol\'ogica de Colombia (UPTC), Tunja 150003, Boyac\'a, Colombia.}
\author{Hanz Y. Ram\'irez}
\email{hanz.ramirez@uptc.edu.co}
\affiliation{Grupo de F\'isica Te\'orica y Computacional, Escuela de F\'isica, Universidad Pedag\'ogica y Tecnol\'ogica de Colombia (UPTC), Tunja 150003, Boyac\'a, Colombia.}

\date{\today}
	
\begin{abstract}
In this theoretical work, we study a double quantum dot interacting strongly with a microcavity, while undergoing resonant tunneling. Effects of interdot tunneling on the light-matter hybridized states are determined, and tunability of their brightness degrees, associated dipole moments, and lifetimes is demonstrated. These results predict dipolariton generation in artificial molecules coupled to optical resonators, and provide a promising scenario for control of emission efficiency and coherence times of exciton polaritons.
\end{abstract}

\maketitle

\section{INTRODUCTION}

In recent years, interest for light generation from low dimensional structures coupled to electrodynamics cavities  has increased noticeably \cite{Kub08,Henn08,kai}. In particular, quantum dots (QDs) have proved to be an excellent tool for experimental observation of purely quantum phenomena, like single photon emission and photon entanglement, both of which can be enhanced through an optical resonator by strengthening the coupling between the QD and the electromagnetic field  \cite{biexciton,purity}.

Those high quality nanostructured semiconductors, that can be obtained by molecular beam epitaxy (MBE) or chemical vapor deposition (CVD) \cite{borovitskaya2002quantum}, have exhibited relevant atom-like phenomena such as Rabi oscillations \cite{ramirez2008coupling,naturetexas}, Mollow triplet in resonance fluorescence \cite{vamivakas2009spin,naturenano}, Dicke effect \cite{vorrath2003dicke,orellana}, and double dressing resonances \cite{hanzRSC,ramirezprl}. Consequently, they are often called ``artificial atoms''. 


Considering the implementation of electronic devices, the use of artificial atoms instead of natural ones would be advantageous, given the obvious convenience of working with stable solid structures rather than with tiny and elusive atoms, for constructing on-chip light-matter hybrid structures \cite{nature2016}.


In turn, microcavities confine light in a small volume and increase radiation-matter coupling as described by the Purcell effect \cite{purcell,lodahl2015interfacing}. In such a strong coupling regime, the system eigenvectors are hybridized states of the QD and the cavity field. These kind of mixed states of light and mater are known as ``exciton polaritons'' (EPs) \cite{coherent,review}. Strong radiation-matter coupling for a QD inside either planar or photonic crystal cavities, has been successfully observed and progressively improved along this century, including electron spin states in charged excitons \cite{vamivakas2009spin,He2013,NatPhot2013,ramirezprl}.

On the other hand, coupling by resonant tunneling between adjacent QDs (artificial molecules) has been proposed as an efficient mechanism to improve tunability in zero dimensional systems \cite{condensed,ramirez2014}. Between different alternatives to control tunneling in double quantum dot (DQD) structures, bias tuning has been found so far as the most successful \cite{hanzbrazil,stinaff,Bracker06}.

For instance, recent experiments have taken advantage of interdot coupling in artificial molecules within QED cavities, for enhancement of hybrid qubits \cite{hybridqubits}, for manipulation of light-matter interaction with superconducting resonators \cite{tunablecavities}, for improvement of single photon emission \cite{patentGammon}, and for observation of phonon assisted cavity feeding  \cite{Petta-phonons}.

Regarding two dimensional systems, several theoretical and experimental efforts have been carried out toward controllable condensation of multiexciton states in double quantum wells inside cavities \cite{double2d-1,double2d-1b,double2d-2,double2d-3,double2d-3b,double2d-5}. Those works revealed the characteristic three branches associated to tunnel coupled structures, which in the 2D case correspond to polariton bands instead of fully discretized polariton states \cite{double2d-4,double2d-4b}. In contrast, reports on dipolaritons in 0D systems are scarce, in despite of their potential usefulness as sources of non-classical light and quantum memories \cite{rojas-unal}. 

As for direct measurement of coherent superpositions of light-matter states in QD-cavity systems, it has been achieved by a few recent experiments \cite{coherent,kai,QDpolariton3,QDpolariton4}. In all cases using individual dots.  

In this work, we study the properties at the small photon number scale of EP modes for a DQD embedded in a microcavity, in such a way that interdot coupling and strong radiation matter interaction are simultaneously considered, and formation of polaritons with adjustable dipole moment (dipolaritons) and reduced brightness (dark polaritons), is explored \cite{dipolaritons}. Hopefully, this investigation may contribute with deeper understanding toward the imminent experimental realization of light-matter hybridized states in DQD-based settings.
 


%
%
%



\section{MODEL}

We consider an asymmetric double quantum dot with a slight difference in the intrinsic energy of the direct and indirect excitonic levels, coupled to a microcavity. Figure 1 a) depicts the proposed system, in which $J$ represents the coupling between left and right dot (LD-RD), while figure 1 b) shows the configurations chosen as basis of the subspace corresponding to the first rung of the Jaynes-Cummings (JC) ladder \cite{cotrino}.

In absence of a bias field, the direct exciton (DX) coupled to a photonic mode would form a conventional polariton with a coupling energy given by the Rabi frequency $\Omega$ (which in turn depends on the radiation-matter constant $g$) \cite{ramirezprl}. On its side, the indirect exciton (IX) is assumed to be a dark state, given the reduced overlap between electron and hole. 

Application of an external bias $F$ on the DQD allows for tuning of the indirect exciton energy, so that resonant tunneling between the $\mid 0,DX \rangle$ and $\mid 0,IX \rangle$ states can be achieved on-demand. 
 
The tunneling rate $J$ depends on the potential barrier experienced by the confined single particles, and is in principle unmodified by the cavity. For simplicity, hole tunneling can be reasonably neglected and then, only electron hopping is considered \cite{Bracker06}. 

\begin{figure}[h]
 \centerline{\includegraphics[scale=0.6]{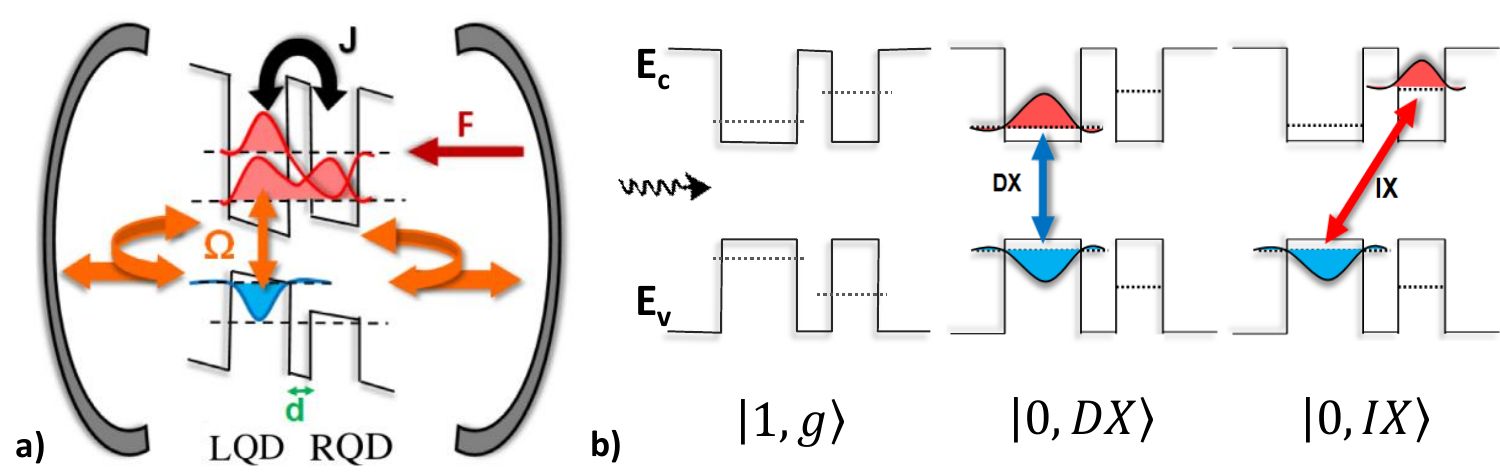}}
 \label{F1}
 \caption{ a) Schematics of the studied system. b) Configuration basis. }
\end{figure}

The Hamiltonian for the $n$-th JC rung in the above described basis reads ($\hbar=1$)

{\small 
\begin{eqnarray}
\hat{H}= & \omega_{C}\hat{n}-\Delta_{c,\,dx}\hat{\sigma}_{dx,g}^{\dagger}\,\hat{\sigma}_{g,dx}^{-}+\left(\Delta_{ix,\,dx}-\Delta_{c,\,dx}-e\,d\,F\right)\hat{\sigma}_{ix,g}^{\dagger}\,\hat{\sigma}_{g,ix}^{-} \nonumber \\
& + g\left(\hat{a}\:\hat{\sigma}_{dx,g}^{\dagger}+\hat{a}^{\dagger}\:\hat{\sigma}_{g,dx}^{-}\right)-\frac{J}{2}\left(\hat{\sigma}_{dx,g}^{\dagger}\hat{\sigma}_{g,ix}^{-}+\hat{\sigma}_{ix,g}^{\dagger}\hat{\sigma}_{g,dx}^{-}\right) \hspace*{1ex},
\end{eqnarray}
}


 
where $\omega_C$ is the cavity mode frequency, $e$ is the electron charge, $d$ is interdot distance (tunneling barrier width), $\Delta_{ix,dx}=\omega_{IX}-\omega_{DX}$ ($\Delta_{c,dx}=\omega_{C}-\omega_{DX}$) is the energy difference between the IX and DX (the cavity mode and the DX), $\hat{n}=\hat{a}^{\dagger}\hat{a}+\hat{\sigma}_{dx,g}^{
\dagger}\,\hat{\sigma}_{g,dx}^{-}+\hat{\sigma}_{ix,g}^{\dagger}\,\hat{\sigma}_{g,ix}^{-}$ is the polariton number operator (with $\hat{a}$ and $\hat{a}^{\dagger}$ the photon annihilation and creation operators, respectively), and $\hat{\sigma}_{dx,g}^{\dagger}=|DX\rangle\langle g|$ ($\hat{\sigma}_{ix,g}^{\dagger}=|IX\rangle\langle g|$) is the transition dipole operator between the DX (IX) and the DQD ground state.  

The control parameter is the electric field $F$, which allows compensating $\Delta_{ix,dx}$ to favor resonant tunneling between dots. Tuning by electrical means is chosen, because as mentioned in the introduction, electric fields have successfully been used as control mechanism in structures coupled by tunneling \cite{stinaff}.

Whether the cavity is a photonic crystal, a micro pillar, or an arrangement of Bragg mirrors, is irrelevant for the model. What becomes important is the existence of a well-defined electromagnetic mode, whose energy difference with the ground direct and indirect excitons be much smaller than the difference between cavity eigenfrequencies. This allows not just neglecting other modes in the cavity, but also the excited levels in each dot constituting the artificial molecule (supposing dots at the order of the few nanometers in height and radius, as the ones used in current quantum optical experiments).    

To obtain the bias dependent radiative lifetimes of the EP eigenstates and their corresponding decay rates, which determine the dynamics of the system at very low temperature, where phonon dissipation effects can be ignored, we use the imaginary part of the effective Hamiltonian yielded by equation (1) 

\begin{equation}
\hat{H}_{eff}=\left(\begin{array}{ccc}
\begin{array}{c}
\omega_{c}n+\Delta_{IX,\,DX}-\Delta_{c,\,DX}-e\,d\,F\:\\
-\frac{i}{2}\left(n-1\right)\kappa\:
\end{array} & -\frac{J}{2}\:\:\:\:\:\:\:\:\:\: & 0\\
-\frac{J}{2}\:\: & \:\begin{array}{c}
\omega_{c}n-\Delta_{c,\,DX}-\frac{i}{2}\gamma_{DX}\\
-\frac{i}{2}\left(n-1\right)\kappa
\end{array}\: & \:\frac{\Omega_{R}}{2}=g\sqrt{n}\\
0 & \:\:\frac{\Omega_{R}}{2}=g\sqrt{n}\: & \:\omega_{c}n-\frac{i}{2}n\kappa
\end{array}\right) \hspace*{1ex},
\end{equation}

in which $\kappa$ represents the cavity scape rate of photons and $\gamma_{DX}$ is the direct exciton recombination rate (we neglect the indirect exciton recombination because of the poor electron-hole overlap in this configuration, i.e. $\gamma_{DX}=0$). 

Thus, the real and imaginary parts of eigenvalues of the matrix in equation (2), respectively provide the EP energies and coherence times of the artificial molecule-cavity system, as functions of the externally applied electric field.   

\section{RESULTS}

By diagonalizing the Hamiltonian in equation (1), the EP modes and their corresponding energies can be obtained ($|1,UP\rangle$, $|1,MP\rangle$ and $|1,LP\rangle$). Figure 2 a) shows the uncoupled and EP energies for the first JC rung as functions of the bias field $F$. Dashed lines describe uncoupled modes (IX-Red, C-Blue, DX green), and the solid lines blue and light blue represent the polariton and dipolariton, respectively. Meanwhile, figure 2 b) shows the fractional components of the basis states (Hopfield coefficients), for each of the EP modes. The following realistic parameters were used in our calculations : $\omega_C=1320.7$ meV, $\Delta_{ix,dx}=80$  mev, $\Delta_{c,dx}=10.7$ meV, $d=15$ nm, $J=0.828$ meV, $g=2\pi 16$ GHz and $\Omega=J$. Under such conditions, the tunneling resonance is found at $-5.75$ kV/cm. This is a remarkably moderate field, very accessible in experiments, in contrast to the high fields necessary for dipolariton manipulation achieved by magnetic means (larger than 5 Tesla), as found by J. S. Rojas-Arias et al. in reference \cite{rojas-unal}.    

\begin{figure}[h]
 \centerline{\includegraphics[scale=0.6]{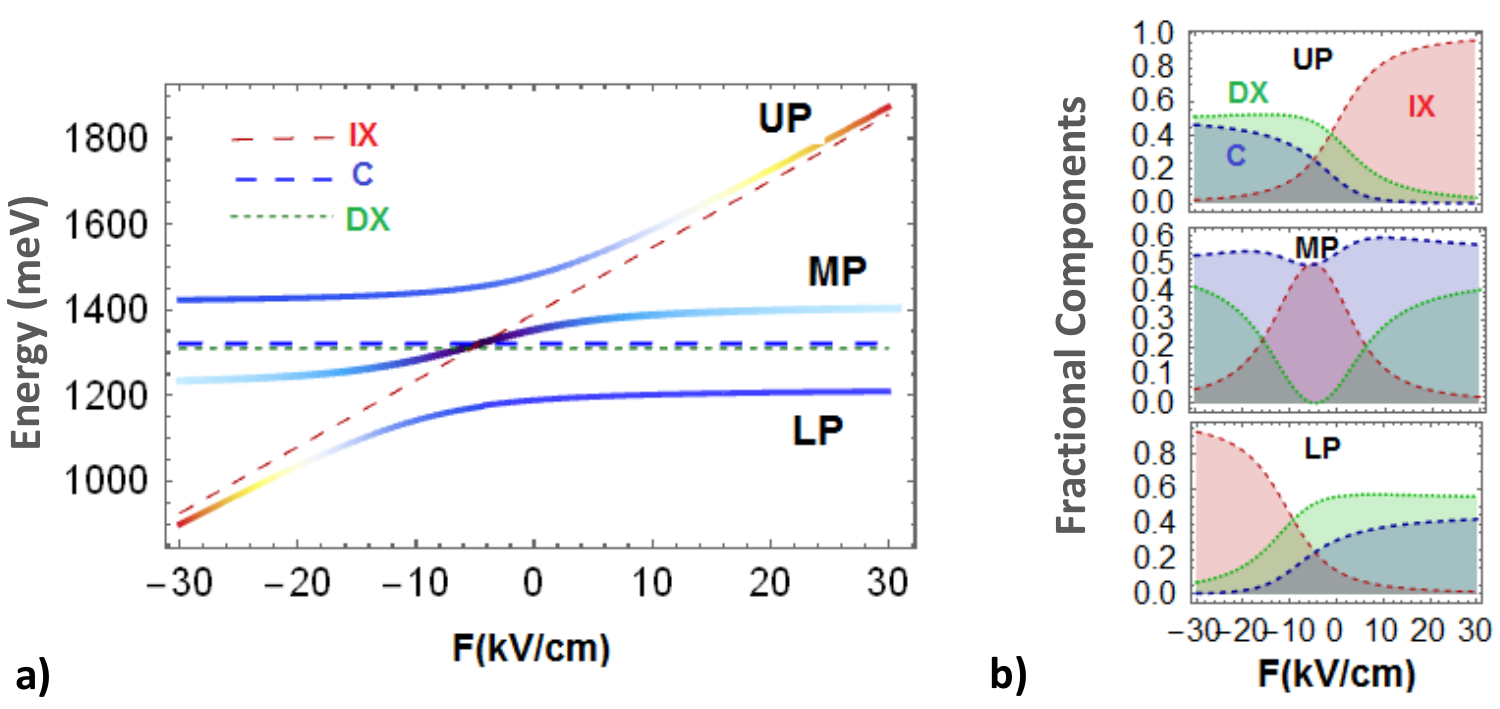}}
  \label{F2}
  \caption{a) Lower, Middle and Upper polariton modes as functions of the bias field $F$. b) Fractional bare components for each of the polariton modes, as functions of $F$. }
 \end{figure}

  

From the obtained coefficients shown in figure 2 b), two associated quantities can be computed to better elucidate the enriched polariton landscape produced by the presence of the second dot. Denoting the EP modes by $\mid 1,LP \rangle , \mid 1,MP \rangle $ and $ \mid 1,UP \rangle $; and considering the superposition

\begin{equation}
\mid 1, \alpha \rangle = C_{1,g}^{\alpha} \mid 1,g \rangle + C_{0,DX}^{\alpha} \mid 0,DX \rangle + C_{0,IX}^{\alpha} \mid 0,IX \rangle \hspace*{1ex},
\end{equation} 
 
where $\alpha=LP,MP,UP$; for each EP mode $\alpha$, we define the bright polariton degree

\begin{equation}
BPD = \mid C_{1,g}^{\alpha} C_{0,DX}^{\alpha} \mid \hspace*{1ex},
\end{equation} 

and the exciton dipole moment 

\begin{equation}
EDM = d \mid C_{0,IX}^{\alpha} \mid \hspace*{1ex}.
\end{equation} 

$BPD$ indicates how strong is the mixing between the DX and the cavity mode, and $EDM$ accounts for the dipole moment associated to the corresponding EP mode. 

Figure 3 shows the $BPD$ and $EDM$ as functions of the bias field $F$, for the same parameters as in figure 2. There, three regimes generated by the interplay between light-matter and interdot coupling, can be observed: (I) Conventional polariton [bright  radiation-matter mixed states with negligible exciton dipole moment], for negative (positive) high values of $F$ in the upper (lower) EP mode, and for positive and negative high values of $F$ in the middle EP mode. (II) Dark dipolariton [mixed radiation matter states with negligible brightness and large dipole moment], for values of $F$ just around the tunneling resonance, in the middle EP mode. (III) Bright dipolariton [mixed radiation matter states with significant brightness and exciton dipole moment], for moderate values of $F$ (as long as they are not very close to the tunneling resonance), in the middle EP mode.    

\begin{figure}[h!]
 \centerline{\includegraphics[scale=0.6]{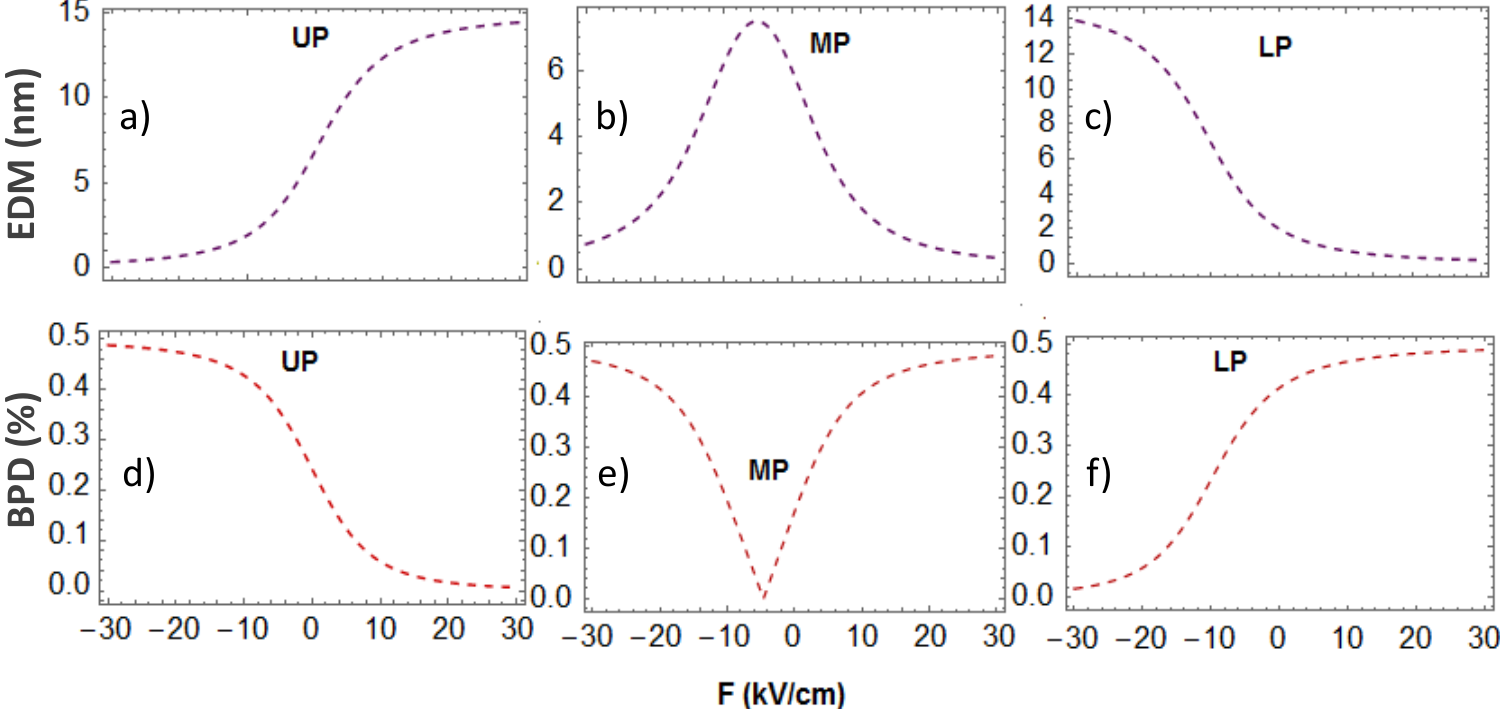}}
  \label{F3}
  \caption{ Top panels: Exciton dipole moments for the a) upper, b) middle and c) lower polariton modes. Bottom panels: Bright polariton degree for the d) upper, e) middle and f) lower polariton modes. both quantities as functions of $F$. }
\end{figure}

Regimes (II) and (III) are particularly interesting. The former because this type of polariton states are expected to be long-living bosons, promising for exciton condensates and derived applications \cite{bose}. The later, because a tunable mixing between bright conventional polaritons and dark dipolaritons provides an optimal scenario for on demand switching between their respective main features.   

The effective tunability by electrical means of the polariton dipole moment (across one order of magnitude), shown in figures 3 a), 3 b) and 3 c), is particularly suggesting and has been not reported in previous related works. 

To evidence how polariton lifetimes can be tuned along a wide range, we calculate the system dynamics by diagonalizing the complex matrix of equation (2) \cite{coherent}. Figure 4 shows the bias field dependence of the recombination rates and lifetimes of all three polaronic branches. The parameters $\gamma_{DX} = 2\pi 0.1$ GHz and $\kappa = 2 \pi 16$ GHz were used in the simulation. 

Those curves reveal how the lower and upper branches allow tuning the polariton lifetimes between tens and hundreds of picoseconds, by application of modest bias (below $|F|<20$ kV/cm). Furthermore, comparing these lifetimes with the ones reported in figures 2 b) and 2 d) of reference \cite{kai}, one can appreciate how there, modulation along similar time ranges require significant temperature fluctuations, which unavoidably imply phonon-related dissipation and dephasing. This evidences improved coherence and stability associated to electric control.           

\begin{figure}[h!]
	\centerline{\includegraphics[scale=1.2]{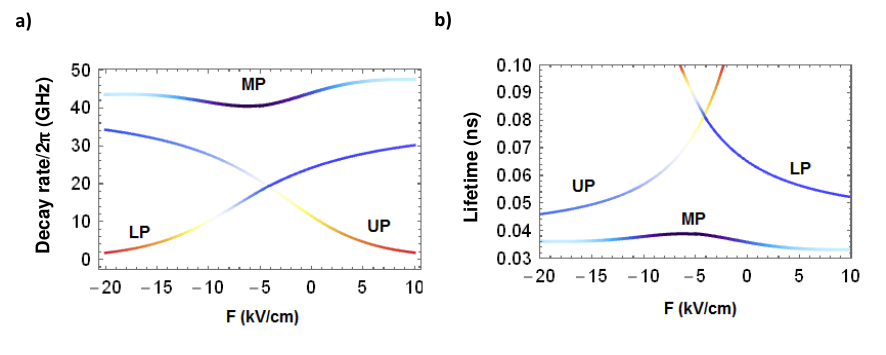}}
	\label{F3}
	\caption{ a) Polariton decay rate and b) polariton lifetime for each EP mode, as function of $F$. }
\end{figure}

\section{CONCLUSION}

In this work, we presented a theoretical model for a quantum dot molecule strongly coupled to a microcavity, which allows calculation of the composed system eigenenergies, as well as of the corresponding eigenstates (dressed states) and radiative decay rates. From the simulated fractional components and polariton lifetimes, as functions of a feasible tuning parameter (electric field), the possibility of generating polaritons with enhanced exciton dipole moment and adjustable emission efficiency and duration is demonstrated.

These results suggest that the proposed combination of artificial molecules with optical resonators, could foster improved control of coherence times and on-demand emission of non-classical light from strongly coupled light-matter arrangements. Thus, further motivation for the experimental realization of exciton polaritons from double quantum dot-cavity settings, is provided.       



\section{ACKNOWLEDGMENTS}

Financial support from the research division of the Universidad Pedag\'ogica y Tecnol\'ogica de Colombia (UPTC), is acknowledged.


\section{CONFLICT OF INTEREST}

The authors declare they have no conflict of interest.



\bibliographystyle{apsrev4-1}
\bibliography{resubmitted}%

\begin{thebibliography}{44}%
\makeatletter
\providecommand \@ifxundefined [1]{%
 \@ifx{#1\undefined}
}%
\providecommand \@ifnum [1]{%
 \ifnum #1\expandafter \@firstoftwo
 \else \expandafter \@secondoftwo
 \fi
}%
\providecommand \@ifx [1]{%
 \ifx #1\expandafter \@firstoftwo
 \else \expandafter \@secondoftwo
 \fi
}%
\providecommand \natexlab [1]{#1}%
\providecommand \enquote  [1]{``#1''}%
\providecommand \bibnamefont  [1]{#1}%
\providecommand \bibfnamefont [1]{#1}%
\providecommand \citenamefont [1]{#1}%
\providecommand \href@noop [0]{\@secondoftwo}%
\providecommand \href [0]{\begingroup \@sanitize@url \@href}%
\providecommand \@href[1]{\@@startlink{#1}\@@href}%
\providecommand \@@href[1]{\endgroup#1\@@endlink}%
\providecommand \@sanitize@url [0]{\catcode `\\12\catcode `\$12\catcode
  `\&12\catcode `\#12\catcode `\^12\catcode `\_12\catcode `\%12\relax}%
\providecommand \@@startlink[1]{}%
\providecommand \@@endlink[0]{}%
\providecommand \url  [0]{\begingroup\@sanitize@url \@url }%
\providecommand \@url [1]{\endgroup\@href {#1}{\urlprefix }}%
\providecommand \urlprefix  [0]{URL }%
\providecommand \Eprint [0]{\href }%
\providecommand \doibase [0]{http://dx.doi.org/}%
\providecommand \selectlanguage [0]{\@gobble}%
\providecommand \bibinfo  [0]{\@secondoftwo}%
\providecommand \bibfield  [0]{\@secondoftwo}%
\providecommand \translation [1]{[#1]}%
\providecommand \BibitemOpen [0]{}%
\providecommand \bibitemStop [0]{}%
\providecommand \bibitemNoStop [0]{.\EOS\space}%
\providecommand \EOS [0]{\spacefactor3000\relax}%
\providecommand \BibitemShut  [1]{\csname bibitem#1\endcsname}%
\let\auto@bib@innerbib\@empty
\bibitem [{\citenamefont {Kubanek}\ \emph {et~al.}(2008)\citenamefont
  {Kubanek}, \citenamefont {Ourjoumtsev}, \citenamefont {Schuster},
  \citenamefont {Koch}, \citenamefont {Pinkse}, \citenamefont {Murr},\ and\
  \citenamefont {Rempe}}]{Kub08}%
  \BibitemOpen
  \bibfield  {author} {\bibinfo {author} {\bibfnamefont {A.}~\bibnamefont
  {Kubanek}}, \bibinfo {author} {\bibfnamefont {A.}~\bibnamefont
  {Ourjoumtsev}}, \bibinfo {author} {\bibfnamefont {I.}~\bibnamefont
  {Schuster}}, \bibinfo {author} {\bibfnamefont {M.}~\bibnamefont {Koch}},
  \bibinfo {author} {\bibfnamefont {P.~W.~H.}\ \bibnamefont {Pinkse}}, \bibinfo
  {author} {\bibfnamefont {K.}~\bibnamefont {Murr}}, \ and\ \bibinfo {author}
  {\bibfnamefont {G.}~\bibnamefont {Rempe}},\ }\href {\doibase
  10.1103/PhysRevLett.101.203602} {\bibfield  {journal} {\bibinfo  {journal}
  {Physical Review Letters}\ }\textbf {\bibinfo {volume} {101}},\ \bibinfo
  {pages} {203602} (\bibinfo {year} {2008})}\BibitemShut {NoStop}%
\bibitem [{\citenamefont {Hennessy}\ \emph {et~al.}(2007)\citenamefont
  {Hennessy}, \citenamefont {Badolato}, \citenamefont {Winger}, \citenamefont
  {Gerace}, \citenamefont {Atature}, \citenamefont {Gulde}, \citenamefont
  {Falt}, \citenamefont {Hu},\ and\ \citenamefont {Imamoglu}}]{Henn08}%
  \BibitemOpen
  \bibfield  {author} {\bibinfo {author} {\bibfnamefont {K.}~\bibnamefont
  {Hennessy}}, \bibinfo {author} {\bibfnamefont {A.}~\bibnamefont {Badolato}},
  \bibinfo {author} {\bibfnamefont {M.}~\bibnamefont {Winger}}, \bibinfo
  {author} {\bibfnamefont {D.}~\bibnamefont {Gerace}}, \bibinfo {author}
  {\bibfnamefont {M.}~\bibnamefont {Atature}}, \bibinfo {author} {\bibfnamefont
  {S.}~\bibnamefont {Gulde}}, \bibinfo {author} {\bibfnamefont
  {S.}~\bibnamefont {Falt}}, \bibinfo {author} {\bibfnamefont {E.~L.}\
  \bibnamefont {Hu}}, \ and\ \bibinfo {author} {\bibfnamefont {A.}~\bibnamefont
  {Imamoglu}},\ }\href {\doibase 10.1038/nature05586} {\bibfield  {journal}
  {\bibinfo  {journal} {Nature Physics}\ }\textbf {\bibinfo {volume} {445}},\
  \bibinfo {pages} {899} (\bibinfo {year} {2007})}\BibitemShut {NoStop}%
\bibitem [{\citenamefont {M\"uller}\ \emph
  {et~al.}(2015{\natexlab{a}})\citenamefont {M\"uller}, \citenamefont
  {Fischer}, \citenamefont {Rundquist}, \citenamefont {Dory}, \citenamefont
  {Lagoudakis}, \citenamefont {Sarmiento}, \citenamefont {Kelaita},
  \citenamefont {Borish},\ and\ \citenamefont {Vu\v{c}kovi\'c}}]{kai}%
  \BibitemOpen
  \bibfield  {author} {\bibinfo {author} {\bibfnamefont {K.}~\bibnamefont
  {M\"uller}}, \bibinfo {author} {\bibfnamefont {K.~A.}\ \bibnamefont
  {Fischer}}, \bibinfo {author} {\bibfnamefont {A.}~\bibnamefont {Rundquist}},
  \bibinfo {author} {\bibfnamefont {C.}~\bibnamefont {Dory}}, \bibinfo {author}
  {\bibfnamefont {K.~G.}\ \bibnamefont {Lagoudakis}}, \bibinfo {author}
  {\bibfnamefont {T.}~\bibnamefont {Sarmiento}}, \bibinfo {author}
  {\bibfnamefont {Y.~A.}\ \bibnamefont {Kelaita}}, \bibinfo {author}
  {\bibfnamefont {V.}~\bibnamefont {Borish}}, \ and\ \bibinfo {author}
  {\bibfnamefont {J.}~\bibnamefont {Vu\v{c}kovi\'c}},\ }\href@noop {}
  {\bibfield  {journal} {\bibinfo  {journal} {Physical Review X}\ }\textbf
  {\bibinfo {volume} {5}},\ \bibinfo {pages} {031006} (\bibinfo {year}
  {2015}{\natexlab{a}})}\BibitemShut {NoStop}%
\bibitem [{\citenamefont {Hargart}\ \emph {et~al.}(2016)\citenamefont
  {Hargart}, \citenamefont {Müller}, \citenamefont {Roy-Choudhury},
  \citenamefont {Portalupi}, \citenamefont {Schneider}, \citenamefont
  {H\"ofling}, \citenamefont {Kamp}, \citenamefont {Hughes},\ and\
  \citenamefont {Michler}}]{biexciton}%
  \BibitemOpen
  \bibfield  {author} {\bibinfo {author} {\bibfnamefont {F.}~\bibnamefont
  {Hargart}}, \bibinfo {author} {\bibfnamefont {M.}~\bibnamefont {Müller}},
  \bibinfo {author} {\bibfnamefont {K.}~\bibnamefont {Roy-Choudhury}}, \bibinfo
  {author} {\bibfnamefont {S.~L.}\ \bibnamefont {Portalupi}}, \bibinfo {author}
  {\bibfnamefont {C.}~\bibnamefont {Schneider}}, \bibinfo {author}
  {\bibfnamefont {S.}~\bibnamefont {H\"ofling}}, \bibinfo {author}
  {\bibfnamefont {M.}~\bibnamefont {Kamp}}, \bibinfo {author} {\bibfnamefont
  {S.}~\bibnamefont {Hughes}}, \ and\ \bibinfo {author} {\bibfnamefont
  {P.}~\bibnamefont {Michler}},\ }\href {\doibase
  http://dx.doi.org/10.1103/PhysRevB.93.115308} {\bibfield  {journal} {\bibinfo
   {journal} {Physical Review B}\ }\textbf {\bibinfo {volume} {93}},\ \bibinfo
  {pages} {115308} (\bibinfo {year} {2016})}\BibitemShut {NoStop}%
\bibitem [{\citenamefont {Xing}\ \emph {et~al.}(2016)\citenamefont {Xing},
  \citenamefont {Yu}, \citenamefont {Duan}, \citenamefont {Gregersen},
  \citenamefont {Chen}, \citenamefont {Unsleber}, \citenamefont {Maier},
  \citenamefont {Schneider}, \citenamefont {Kamp}, \citenamefont {H\"ofling},
  \citenamefont {Lu},\ and\ \citenamefont {Pan}}]{purity}%
  \BibitemOpen
  \bibfield  {author} {\bibinfo {author} {\bibfnamefont {D.}~\bibnamefont
  {Xing}}, \bibinfo {author} {\bibfnamefont {H.}~\bibnamefont {Yu}}, \bibinfo
  {author} {\bibfnamefont {Z.~C.}\ \bibnamefont {Duan}}, \bibinfo {author}
  {\bibfnamefont {N.}~\bibnamefont {Gregersen}}, \bibinfo {author}
  {\bibfnamefont {M.~C.}\ \bibnamefont {Chen}}, \bibinfo {author}
  {\bibfnamefont {S.}~\bibnamefont {Unsleber}}, \bibinfo {author}
  {\bibfnamefont {S.}~\bibnamefont {Maier}}, \bibinfo {author} {\bibfnamefont
  {C.}~\bibnamefont {Schneider}}, \bibinfo {author} {\bibfnamefont
  {M.}~\bibnamefont {Kamp}}, \bibinfo {author} {\bibfnamefont {S.}~\bibnamefont
  {H\"ofling}}, \bibinfo {author} {\bibfnamefont {C.~Y.}\ \bibnamefont {Lu}}, \
  and\ \bibinfo {author} {\bibfnamefont {J.~W.}\ \bibnamefont {Pan}},\ }\href
  {\doibase http://dx.doi.org/10.1103/PhysRevLett.116.020401} {\bibfield
  {journal} {\bibinfo  {journal} {Physical Review Letters}\ }\textbf {\bibinfo
  {volume} {116}},\ \bibinfo {pages} {020401} (\bibinfo {year}
  {2016})}\BibitemShut {NoStop}%
\bibitem [{\citenamefont {Borovitskaya}\ and\ \citenamefont
  {Shur}(2002)}]{borovitskaya2002quantum}%
  \BibitemOpen
  \bibfield  {author} {\bibinfo {author} {\bibfnamefont {E.}~\bibnamefont
  {Borovitskaya}}\ and\ \bibinfo {author} {\bibfnamefont {M.~S.}\ \bibnamefont
  {Shur}},\ }\href@noop {} {\emph {\bibinfo {title} {Quantum dots}}},\
  Vol.~\bibinfo {volume} {25}\ (\bibinfo  {publisher} {World Scientific},\
  \bibinfo {year} {2002})\BibitemShut {NoStop}%
\bibitem [{\citenamefont {Ram\'irez}\ and\ \citenamefont
  {Camacho}(2008)}]{ramirez2008coupling}%
  \BibitemOpen
  \bibfield  {author} {\bibinfo {author} {\bibfnamefont {H.~Y.}\ \bibnamefont
  {Ram\'irez}}\ and\ \bibinfo {author} {\bibfnamefont {A.~S.}\ \bibnamefont
  {Camacho}},\ }\href {\doibase https://doi.org/10.1016/j.physe.2008.02.009}
  {\bibfield  {journal} {\bibinfo  {journal} {Physica E}\ }\textbf {\bibinfo
  {volume} {40}},\ \bibinfo {pages} {2937} (\bibinfo {year}
  {2008})}\BibitemShut {NoStop}%
\bibitem [{\citenamefont {Flagg}\ \emph {et~al.}(2009)\citenamefont {Flagg},
  \citenamefont {Muller}, \citenamefont {Robertson}, \citenamefont {Founta},
  \citenamefont {Deppe}, \citenamefont {Xiao}, \citenamefont {Ma},
  \citenamefont {Salamo},\ and\ \citenamefont {Shih}}]{naturetexas}%
  \BibitemOpen
  \bibfield  {author} {\bibinfo {author} {\bibfnamefont {E.~B.}\ \bibnamefont
  {Flagg}}, \bibinfo {author} {\bibfnamefont {A.}~\bibnamefont {Muller}},
  \bibinfo {author} {\bibfnamefont {J.~W.}\ \bibnamefont {Robertson}}, \bibinfo
  {author} {\bibfnamefont {S.}~\bibnamefont {Founta}}, \bibinfo {author}
  {\bibfnamefont {D.~G.}\ \bibnamefont {Deppe}}, \bibinfo {author}
  {\bibfnamefont {M.}~\bibnamefont {Xiao}}, \bibinfo {author} {\bibfnamefont
  {W.}~\bibnamefont {Ma}}, \bibinfo {author} {\bibfnamefont {G.~J.}\
  \bibnamefont {Salamo}}, \ and\ \bibinfo {author} {\bibfnamefont {C.~K.}\
  \bibnamefont {Shih}},\ }\href {\doibase https://doi.org/10.1038/nphys1184}
  {\bibfield  {journal} {\bibinfo  {journal} {Nature Physics}\ }\textbf
  {\bibinfo {volume} {5}},\ \bibinfo {pages} {203} (\bibinfo {year}
  {2009})}\BibitemShut {NoStop}%
\bibitem [{\citenamefont {Vamivakas}\ \emph {et~al.}(2009)\citenamefont
  {Vamivakas}, \citenamefont {Zhao}, \citenamefont {Lu},\ and\ \citenamefont
  {Atat{\"u}re}}]{vamivakas2009spin}%
  \BibitemOpen
  \bibfield  {author} {\bibinfo {author} {\bibfnamefont {A.~N.}\ \bibnamefont
  {Vamivakas}}, \bibinfo {author} {\bibfnamefont {Y.}~\bibnamefont {Zhao}},
  \bibinfo {author} {\bibfnamefont {C.-Y.}\ \bibnamefont {Lu}}, \ and\ \bibinfo
  {author} {\bibfnamefont {M.}~\bibnamefont {Atat{\"u}re}},\ }\href@noop {}
  {\bibfield  {journal} {\bibinfo  {journal} {Nature Physics}\ }\textbf
  {\bibinfo {volume} {5}},\ \bibinfo {pages} {198} (\bibinfo {year}
  {2009})}\BibitemShut {NoStop}%
\bibitem [{\citenamefont {He}\ \emph {et~al.}(2013{\natexlab{a}})\citenamefont
  {He}, \citenamefont {He}, \citenamefont {Wei}, \citenamefont {Wu},
  \citenamefont {Atatüre}, \citenamefont {Schneider}, \citenamefont
  {Höfling}, \citenamefont {Kamp}, \citenamefont {Lu},\ and\ \citenamefont
  {Pan}}]{naturenano}%
  \BibitemOpen
  \bibfield  {author} {\bibinfo {author} {\bibfnamefont {Y.~M.}\ \bibnamefont
  {He}}, \bibinfo {author} {\bibfnamefont {Y.}~\bibnamefont {He}}, \bibinfo
  {author} {\bibfnamefont {Y.~J.}\ \bibnamefont {Wei}}, \bibinfo {author}
  {\bibfnamefont {D.}~\bibnamefont {Wu}}, \bibinfo {author} {\bibfnamefont
  {M.}~\bibnamefont {Atatüre}}, \bibinfo {author} {\bibfnamefont
  {C.}~\bibnamefont {Schneider}}, \bibinfo {author} {\bibfnamefont
  {S.}~\bibnamefont {Höfling}}, \bibinfo {author} {\bibfnamefont
  {M.}~\bibnamefont {Kamp}}, \bibinfo {author} {\bibfnamefont {C.~Y.}\
  \bibnamefont {Lu}}, \ and\ \bibinfo {author} {\bibfnamefont {J.~W.}\
  \bibnamefont {Pan}},\ }\href {\doibase
  https://doi.org/10.1038/nnano.2012.262} {\bibfield  {journal} {\bibinfo
  {journal} {Nature Nanotechnology}\ }\textbf {\bibinfo {volume} {8}},\
  \bibinfo {pages} {213} (\bibinfo {year} {2013}{\natexlab{a}})}\BibitemShut
  {NoStop}%
\bibitem [{\citenamefont {Vorrath}\ and\ \citenamefont
  {Brandes}(2003)}]{vorrath2003dicke}%
  \BibitemOpen
  \bibfield  {author} {\bibinfo {author} {\bibfnamefont {T.}~\bibnamefont
  {Vorrath}}\ and\ \bibinfo {author} {\bibfnamefont {T.}~\bibnamefont
  {Brandes}},\ }\href@noop {} {\bibfield  {journal} {\bibinfo  {journal}
  {Physical Review B}\ }\textbf {\bibinfo {volume} {68}},\ \bibinfo {pages}
  {035309} (\bibinfo {year} {2003})}\BibitemShut {NoStop}%
\bibitem [{\citenamefont {Orellana}\ \emph {et~al.}(2006)\citenamefont
  {Orellana}, \citenamefont {Lara},\ and\ \citenamefont {V}}]{orellana}%
  \BibitemOpen
  \bibfield  {author} {\bibinfo {author} {\bibfnamefont {P.~A.}\ \bibnamefont
  {Orellana}}, \bibinfo {author} {\bibfnamefont {G.~A.}\ \bibnamefont {Lara}},
  \ and\ \bibinfo {author} {\bibfnamefont {A.~E.}\ \bibnamefont {V}},\ }\href
  {\doibase https://doi.org/10.1103/PhysRevB.74.193315} {\bibfield  {journal}
  {\bibinfo  {journal} {Physical Review B}\ }\textbf {\bibinfo {volume} {74}},\
  \bibinfo {pages} {193315} (\bibinfo {year} {2006})}\BibitemShut {NoStop}%
\bibitem [{\citenamefont {Ram\'irez}(2013)}]{hanzRSC}%
  \BibitemOpen
  \bibfield  {author} {\bibinfo {author} {\bibfnamefont {H.~Y.}\ \bibnamefont
  {Ram\'irez}},\ }\href {\doibase 10.1039/c3ra43749c} {\bibfield  {journal}
  {\bibinfo  {journal} {RSC Advances}\ }\textbf {\bibinfo {volume} {3}},\
  \bibinfo {pages} {24991} (\bibinfo {year} {2013})}\BibitemShut {NoStop}%
\bibitem [{\citenamefont {He}\ \emph {et~al.}(2015)\citenamefont {He},
  \citenamefont {He}, \citenamefont {Liu}, \citenamefont {Wei}, \citenamefont
  {Ram\'irez}, \citenamefont {Atat\"ure}, \citenamefont {Schneider},
  \citenamefont {H\"ofling}, \citenamefont {Lu},\ and\ \citenamefont
  {Pan}}]{ramirezprl}%
  \BibitemOpen
  \bibfield  {author} {\bibinfo {author} {\bibfnamefont {Y.}~\bibnamefont
  {He}}, \bibinfo {author} {\bibfnamefont {Y.~M.}\ \bibnamefont {He}}, \bibinfo
  {author} {\bibfnamefont {J.}~\bibnamefont {Liu}}, \bibinfo {author}
  {\bibfnamefont {Y.~J.}\ \bibnamefont {Wei}}, \bibinfo {author} {\bibfnamefont
  {H.~Y.}\ \bibnamefont {Ram\'irez}}, \bibinfo {author} {\bibfnamefont
  {M.}~\bibnamefont {Atat\"ure}}, \bibinfo {author} {\bibfnamefont
  {C.}~\bibnamefont {Schneider}}, \bibinfo {author} {\bibfnamefont
  {S.}~\bibnamefont {H\"ofling}}, \bibinfo {author} {\bibfnamefont {C.~Y.}\
  \bibnamefont {Lu}}, \ and\ \bibinfo {author} {\bibfnamefont {J.~W.}\
  \bibnamefont {Pan}},\ }\href@noop {} {\bibfield  {journal} {\bibinfo
  {journal} {Physical Review Letters}\ }\textbf {\bibinfo {volume} {114}},\
  \bibinfo {pages} {097402} (\bibinfo {year} {2015})}\BibitemShut {NoStop}%
\bibitem [{\citenamefont {Aharonovich}\ \emph {et~al.}(2016)\citenamefont
  {Aharonovich}, \citenamefont {Englund},\ and\ \citenamefont
  {Toth}}]{nature2016}%
  \BibitemOpen
  \bibfield  {author} {\bibinfo {author} {\bibfnamefont {I.}~\bibnamefont
  {Aharonovich}}, \bibinfo {author} {\bibfnamefont {D.}~\bibnamefont
  {Englund}}, \ and\ \bibinfo {author} {\bibfnamefont {M.}~\bibnamefont
  {Toth}},\ }\href {\doibase https://doi.org/10.1038/nphoton.2016.186}
  {\bibfield  {journal} {\bibinfo  {journal} {Nature Photonics}\ }\textbf
  {\bibinfo {volume} {10}},\ \bibinfo {pages} {631} (\bibinfo {year}
  {2016})}\BibitemShut {NoStop}%
\bibitem [{\citenamefont {Purcell}(1946)}]{purcell}%
  \BibitemOpen
  \bibfield  {author} {\bibinfo {author} {\bibfnamefont {E.~M.}\ \bibnamefont
  {Purcell}},\ }\href@noop {} {\bibfield  {journal} {\bibinfo  {journal}
  {physical Review}\ }\textbf {\bibinfo {volume} {69}},\ \bibinfo {pages} {681}
  (\bibinfo {year} {1946})}\BibitemShut {NoStop}%
\bibitem [{\citenamefont {Lodahl}\ \emph {et~al.}(2015)\citenamefont {Lodahl},
  \citenamefont {Mahmoodian},\ and\ \citenamefont
  {Stobbe}}]{lodahl2015interfacing}%
  \BibitemOpen
  \bibfield  {author} {\bibinfo {author} {\bibfnamefont {P.}~\bibnamefont
  {Lodahl}}, \bibinfo {author} {\bibfnamefont {S.}~\bibnamefont {Mahmoodian}},
  \ and\ \bibinfo {author} {\bibfnamefont {S.}~\bibnamefont {Stobbe}},\
  }\href@noop {} {\bibfield  {journal} {\bibinfo  {journal} {Reviews of Modern
  Physics}\ }\textbf {\bibinfo {volume} {87}},\ \bibinfo {pages} {347}
  (\bibinfo {year} {2015})}\BibitemShut {NoStop}%
\bibitem [{\citenamefont {M\"uller}\ \emph
  {et~al.}(2015{\natexlab{b}})\citenamefont {M\"uller}, \citenamefont
  {Rundquist}, \citenamefont {Fischer}, \citenamefont {Sarmiento},
  \citenamefont {Lagoudakis}, \citenamefont {Kelaita}, \citenamefont
  {S\'anchez-Mu\'noz}, \citenamefont {del Valle}, \citenamefont {Laussy},\ and\
  \citenamefont {Vu\v{c}kovi\'c}}]{coherent}%
  \BibitemOpen
  \bibfield  {author} {\bibinfo {author} {\bibfnamefont {K.}~\bibnamefont
  {M\"uller}}, \bibinfo {author} {\bibfnamefont {A.}~\bibnamefont {Rundquist}},
  \bibinfo {author} {\bibfnamefont {K.~A.}\ \bibnamefont {Fischer}}, \bibinfo
  {author} {\bibfnamefont {T.}~\bibnamefont {Sarmiento}}, \bibinfo {author}
  {\bibfnamefont {K.~G.}\ \bibnamefont {Lagoudakis}}, \bibinfo {author}
  {\bibfnamefont {Y.}~\bibnamefont {Kelaita}}, \bibinfo {author} {\bibfnamefont
  {A.~C.}\ \bibnamefont {S\'anchez-Mu\'noz}}, \bibinfo {author} {\bibfnamefont
  {E.}~\bibnamefont {del Valle}}, \bibinfo {author} {\bibfnamefont {F.~P.}\
  \bibnamefont {Laussy}}, \ and\ \bibinfo {author} {\bibfnamefont
  {J.}~\bibnamefont {Vu\v{c}kovi\'c}},\ }\href@noop {} {\bibfield  {journal}
  {\bibinfo  {journal} {Physical Review Letters}\ }\textbf {\bibinfo {volume}
  {114}},\ \bibinfo {pages} {233601} (\bibinfo {year}
  {2015}{\natexlab{b}})}\BibitemShut {NoStop}%
\bibitem [{\citenamefont {Schneider}\ \emph {et~al.}(2017)\citenamefont
  {Schneider}, \citenamefont {Winkler}, \citenamefont {Fraser}, \citenamefont
  {Kamp}, \citenamefont {Yamamoto}, \citenamefont {Ostrovskaya},\ and\
  \citenamefont {H\"ofling}}]{review}%
  \BibitemOpen
  \bibfield  {author} {\bibinfo {author} {\bibfnamefont {C.}~\bibnamefont
  {Schneider}}, \bibinfo {author} {\bibfnamefont {K.}~\bibnamefont {Winkler}},
  \bibinfo {author} {\bibfnamefont {M.~D.}\ \bibnamefont {Fraser}}, \bibinfo
  {author} {\bibfnamefont {M.}~\bibnamefont {Kamp}}, \bibinfo {author}
  {\bibfnamefont {Y.}~\bibnamefont {Yamamoto}}, \bibinfo {author}
  {\bibfnamefont {E.~A.}\ \bibnamefont {Ostrovskaya}}, \ and\ \bibinfo {author}
  {\bibfnamefont {S.}~\bibnamefont {H\"ofling}},\ }\href@noop {} {\bibfield
  {journal} {\bibinfo  {journal} {Reports on Progress in Physics}\ }\textbf
  {\bibinfo {volume} {80}},\ \bibinfo {pages} {016503} (\bibinfo {year}
  {2017})}\BibitemShut {NoStop}%
\bibitem [{\citenamefont {He}\ \emph {et~al.}(2013{\natexlab{b}})\citenamefont
  {He}, \citenamefont {He}, \citenamefont {Wei}, \citenamefont {Wu},
  \citenamefont {Atat\"ure}, \citenamefont {Schneider}, \citenamefont
  {H\"ofling}, \citenamefont {Kamp}, \citenamefont {Lu},\ and\ \citenamefont
  {Pan}}]{He2013}%
  \BibitemOpen
  \bibfield  {author} {\bibinfo {author} {\bibfnamefont {Y.~M.}\ \bibnamefont
  {He}}, \bibinfo {author} {\bibfnamefont {Y.}~\bibnamefont {He}}, \bibinfo
  {author} {\bibfnamefont {Y.~J.}\ \bibnamefont {Wei}}, \bibinfo {author}
  {\bibfnamefont {D.}~\bibnamefont {Wu}}, \bibinfo {author} {\bibfnamefont
  {M.}~\bibnamefont {Atat\"ure}}, \bibinfo {author} {\bibfnamefont
  {C.}~\bibnamefont {Schneider}}, \bibinfo {author} {\bibfnamefont
  {S.}~\bibnamefont {H\"ofling}}, \bibinfo {author} {\bibfnamefont
  {M.}~\bibnamefont {Kamp}}, \bibinfo {author} {\bibfnamefont {C.~Y.}\
  \bibnamefont {Lu}}, \ and\ \bibinfo {author} {\bibfnamefont {J.~W.}\
  \bibnamefont {Pan}},\ }\href@noop {} {\bibfield  {journal} {\bibinfo
  {journal} {Nature Nanotechnology}\ }\textbf {\bibinfo {volume} {8}},\
  \bibinfo {pages} {213} (\bibinfo {year} {2013}{\natexlab{b}})}\BibitemShut
  {NoStop}%
\bibitem [{\citenamefont {Carter}\ \emph {et~al.}(2013)\citenamefont {Carter},
  \citenamefont {Sweeney}, \citenamefont {Kim}, \citenamefont {Kim},
  \citenamefont {Solenov}, \citenamefont {Economou}, \citenamefont {Reinecke},
  \citenamefont {Yang}, \citenamefont {Bracker},\ and\ \citenamefont
  {Gammon}}]{NatPhot2013}%
  \BibitemOpen
  \bibfield  {author} {\bibinfo {author} {\bibfnamefont {S.~G.}\ \bibnamefont
  {Carter}}, \bibinfo {author} {\bibfnamefont {T.~M.}\ \bibnamefont {Sweeney}},
  \bibinfo {author} {\bibfnamefont {M.~J.}\ \bibnamefont {Kim}}, \bibinfo
  {author} {\bibfnamefont {C.~S.}\ \bibnamefont {Kim}}, \bibinfo {author}
  {\bibfnamefont {D.}~\bibnamefont {Solenov}}, \bibinfo {author} {\bibfnamefont
  {S.~E.}\ \bibnamefont {Economou}}, \bibinfo {author} {\bibfnamefont {T.~L.}\
  \bibnamefont {Reinecke}}, \bibinfo {author} {\bibfnamefont {L.}~\bibnamefont
  {Yang}}, \bibinfo {author} {\bibfnamefont {A.~S.}\ \bibnamefont {Bracker}}, \
  and\ \bibinfo {author} {\bibfnamefont {D.}~\bibnamefont {Gammon}},\
  }\href@noop {} {\bibfield  {journal} {\bibinfo  {journal} {Nature Photonics}\
  }\textbf {\bibinfo {volume} {7}},\ \bibinfo {pages} {329} (\bibinfo {year}
  {2013})}\BibitemShut {NoStop}%
\bibitem [{\citenamefont {Ram\'irez}\ \emph {et~al.}(2007)\citenamefont
  {Ram\'irez}, \citenamefont {Camacho},\ and\ \citenamefont {Lew
  Yan~Voon}}]{condensed}%
  \BibitemOpen
  \bibfield  {author} {\bibinfo {author} {\bibfnamefont {H.~Y.}\ \bibnamefont
  {Ram\'irez}}, \bibinfo {author} {\bibfnamefont {A.~S.}\ \bibnamefont
  {Camacho}}, \ and\ \bibinfo {author} {\bibfnamefont {L.~C.}\ \bibnamefont
  {Lew Yan~Voon}},\ }\href@noop {} {\bibfield  {journal} {\bibinfo  {journal}
  {Journal of Physics: Condensed Matter}\ }\textbf {\bibinfo {volume} {19}},\
  \bibinfo {pages} {346216} (\bibinfo {year} {2007})}\BibitemShut {NoStop}%
\bibitem [{\citenamefont {Fino}\ \emph {et~al.}(2014)\citenamefont {Fino},
  \citenamefont {Camacho},\ and\ \citenamefont {Ram\'irez}}]{ramirez2014}%
  \BibitemOpen
  \bibfield  {author} {\bibinfo {author} {\bibfnamefont {N.~R.}\ \bibnamefont
  {Fino}}, \bibinfo {author} {\bibfnamefont {A.~S.}\ \bibnamefont {Camacho}}, \
  and\ \bibinfo {author} {\bibfnamefont {H.~Y.}\ \bibnamefont {Ram\'irez}},\
  }\href@noop {} {\bibfield  {journal} {\bibinfo  {journal} {Nanoscale Research
  Letters}\ }\textbf {\bibinfo {volume} {9}},\ \bibinfo {pages} {297} (\bibinfo
  {year} {2014})}\BibitemShut {NoStop}%
\bibitem [{\citenamefont {Ram\'irez}\ \emph {et~al.}(2006)\citenamefont
  {Ram\'irez}, \citenamefont {Camacho},\ and\ \citenamefont {Lew
  Yan~Voon}}]{hanzbrazil}%
  \BibitemOpen
  \bibfield  {author} {\bibinfo {author} {\bibfnamefont {H.~Y.}\ \bibnamefont
  {Ram\'irez}}, \bibinfo {author} {\bibfnamefont {A.~S.}\ \bibnamefont
  {Camacho}}, \ and\ \bibinfo {author} {\bibfnamefont {L.~C.}\ \bibnamefont
  {Lew Yan~Voon}},\ }\href@noop {} {\bibfield  {journal} {\bibinfo  {journal}
  {Brazilian Journal of Physics}\ }\textbf {\bibinfo {volume} {36}},\ \bibinfo
  {pages} {869} (\bibinfo {year} {2006})}\BibitemShut {NoStop}%
\bibitem [{\citenamefont {Stinaff}\ \emph {et~al.}(2006)\citenamefont
  {Stinaff}, \citenamefont {Scheibner}, \citenamefont {Bracker}, \citenamefont
  {Ponomarev}, \citenamefont {Korenev}, \citenamefont {Ware}, \citenamefont
  {Doty}, \citenamefont {Reinecke},\ and\ \citenamefont {Gammon}}]{stinaff}%
  \BibitemOpen
  \bibfield  {author} {\bibinfo {author} {\bibfnamefont {E.~A.}\ \bibnamefont
  {Stinaff}}, \bibinfo {author} {\bibfnamefont {M.}~\bibnamefont {Scheibner}},
  \bibinfo {author} {\bibfnamefont {A.~S.}\ \bibnamefont {Bracker}}, \bibinfo
  {author} {\bibfnamefont {I.~V.}\ \bibnamefont {Ponomarev}}, \bibinfo {author}
  {\bibfnamefont {V.~L.}\ \bibnamefont {Korenev}}, \bibinfo {author}
  {\bibfnamefont {M.~E.}\ \bibnamefont {Ware}}, \bibinfo {author}
  {\bibfnamefont {M.~F.}\ \bibnamefont {Doty}}, \bibinfo {author}
  {\bibfnamefont {T.~L.}\ \bibnamefont {Reinecke}}, \ and\ \bibinfo {author}
  {\bibfnamefont {D.}~\bibnamefont {Gammon}},\ }\href@noop {} {\bibfield
  {journal} {\bibinfo  {journal} {Science}\ }\textbf {\bibinfo {volume}
  {311}},\ \bibinfo {pages} {636} (\bibinfo {year} {2006})}\BibitemShut
  {NoStop}%
\bibitem [{\citenamefont {Bracker}\ \emph {et~al.}(2006)\citenamefont
  {Bracker}, \citenamefont {Scheibner}, \citenamefont {Doty}, \citenamefont
  {Stinaff}, \citenamefont {Ponomarev}, \citenamefont {Kim}, \citenamefont
  {Whitman}, \citenamefont {Reinecke},\ and\ \citenamefont
  {Gammon}}]{Bracker06}%
  \BibitemOpen
  \bibfield  {author} {\bibinfo {author} {\bibfnamefont {A.~S.}\ \bibnamefont
  {Bracker}}, \bibinfo {author} {\bibfnamefont {M.}~\bibnamefont {Scheibner}},
  \bibinfo {author} {\bibfnamefont {M.~F.}\ \bibnamefont {Doty}}, \bibinfo
  {author} {\bibfnamefont {E.~A.}\ \bibnamefont {Stinaff}}, \bibinfo {author}
  {\bibfnamefont {I.~V.}\ \bibnamefont {Ponomarev}}, \bibinfo {author}
  {\bibfnamefont {J.~C.}\ \bibnamefont {Kim}}, \bibinfo {author} {\bibfnamefont
  {L.~J.}\ \bibnamefont {Whitman}}, \bibinfo {author} {\bibfnamefont {T.~L.}\
  \bibnamefont {Reinecke}}, \ and\ \bibinfo {author} {\bibfnamefont
  {D.}~\bibnamefont {Gammon}},\ }\href {\doibase
  http://dx.doi.org/10.1063/1.2400397} {\bibfield  {journal} {\bibinfo
  {journal} {Applied Physics Letters}\ }\textbf {\bibinfo {volume} {89}},\
  \bibinfo {pages} {23} (\bibinfo {year} {2006})}\BibitemShut {NoStop}%
\bibitem [{\citenamefont {Vora}\ \emph {et~al.}(2015)\citenamefont {Vora},
  \citenamefont {Bracker}, \citenamefont {Carter}, \citenamefont {Sweeney},
  \citenamefont {Kim}, \citenamefont {Kim}, \citenamefont {Yang}, \citenamefont
  {Brereton}, \citenamefont {E},\ and\ \citenamefont {D}}]{hybridqubits}%
  \BibitemOpen
  \bibfield  {author} {\bibinfo {author} {\bibfnamefont {P.~M.}\ \bibnamefont
  {Vora}}, \bibinfo {author} {\bibfnamefont {A.~S.}\ \bibnamefont {Bracker}},
  \bibinfo {author} {\bibfnamefont {S.~G.}\ \bibnamefont {Carter}}, \bibinfo
  {author} {\bibfnamefont {T.~M.}\ \bibnamefont {Sweeney}}, \bibinfo {author}
  {\bibfnamefont {M.~J.}\ \bibnamefont {Kim}}, \bibinfo {author} {\bibfnamefont
  {C.~S.}\ \bibnamefont {Kim}}, \bibinfo {author} {\bibfnamefont
  {L.}~\bibnamefont {Yang}}, \bibinfo {author} {\bibfnamefont {P.~G.}\
  \bibnamefont {Brereton}}, \bibinfo {author} {\bibfnamefont {E.~S.}\
  \bibnamefont {E}}, \ and\ \bibinfo {author} {\bibfnamefont {G.}~\bibnamefont
  {D}},\ }\href {\doibase doi: 10.1038/ncomms8665 (2015)} {\bibfield  {journal}
  {\bibinfo  {journal} {Nature Communications}\ }\textbf {\bibinfo {volume}
  {6}},\ \bibinfo {pages} {7665} (\bibinfo {year} {2015})}\BibitemShut
  {NoStop}%
\bibitem [{\citenamefont {Stockklauser}\ \emph {et~al.}(2017)\citenamefont
  {Stockklauser}, \citenamefont {Scarlino}, \citenamefont {Koski},
  \citenamefont {Gasparinetti}, \citenamefont {Andersen}, \citenamefont
  {Reichl}, \citenamefont {Wegscheider}, \citenamefont {Ihn}, \citenamefont
  {Ensslin},\ and\ \citenamefont {Wallraff}}]{tunablecavities}%
  \BibitemOpen
  \bibfield  {author} {\bibinfo {author} {\bibfnamefont {A.}~\bibnamefont
  {Stockklauser}}, \bibinfo {author} {\bibfnamefont {P.}~\bibnamefont
  {Scarlino}}, \bibinfo {author} {\bibfnamefont {J.~V.}\ \bibnamefont {Koski}},
  \bibinfo {author} {\bibfnamefont {S.}~\bibnamefont {Gasparinetti}}, \bibinfo
  {author} {\bibfnamefont {C.~K.}\ \bibnamefont {Andersen}}, \bibinfo {author}
  {\bibfnamefont {C.}~\bibnamefont {Reichl}}, \bibinfo {author} {\bibfnamefont
  {W.}~\bibnamefont {Wegscheider}}, \bibinfo {author} {\bibfnamefont
  {T.}~\bibnamefont {Ihn}}, \bibinfo {author} {\bibfnamefont {K.}~\bibnamefont
  {Ensslin}}, \ and\ \bibinfo {author} {\bibfnamefont {A.}~\bibnamefont
  {Wallraff}},\ }\href
  {https://journals.aps.org/prx/abstract/10.1103/PhysRevX.7.011030} {\bibfield
  {journal} {\bibinfo  {journal} {Physical Review X}\ }\textbf {\bibinfo
  {volume} {7}},\ \bibinfo {pages} {011030} (\bibinfo {year}
  {2017})}\BibitemShut {NoStop}%
\bibitem [{\citenamefont {Gammon}\ \emph {et~al.}(2017)\citenamefont {Gammon},
  \citenamefont {Carter}, \citenamefont {Bracker},\ and\ \citenamefont
  {Vora}}]{patentGammon}%
  \BibitemOpen
  \bibfield  {author} {\bibinfo {author} {\bibfnamefont {D.}~\bibnamefont
  {Gammon}}, \bibinfo {author} {\bibfnamefont {S.}~\bibnamefont {Carter}},
  \bibinfo {author} {\bibfnamefont {A.~S.}\ \bibnamefont {Bracker}}, \ and\
  \bibinfo {author} {\bibfnamefont {P.}~\bibnamefont {Vora}},\ }\href
  {https://patents.google.com/patent/US9619754} {\bibfield  {journal} {\bibinfo
   {journal} {US Patent}\ }\textbf {\bibinfo {volume} {US9619754B2}} (\bibinfo
  {year} {2017})}\BibitemShut {NoStop}%
\bibitem [{\citenamefont {Hartke}\ \emph {et~al.}(2018)\citenamefont {Hartke},
  \citenamefont {Liu}, \citenamefont {Gullans},\ and\ \citenamefont
  {R}}]{Petta-phonons}%
  \BibitemOpen
  \bibfield  {author} {\bibinfo {author} {\bibfnamefont {T.~R.}\ \bibnamefont
  {Hartke}}, \bibinfo {author} {\bibfnamefont {Y.~Y.}\ \bibnamefont {Liu}},
  \bibinfo {author} {\bibfnamefont {M.~J.}\ \bibnamefont {Gullans}}, \ and\
  \bibinfo {author} {\bibfnamefont {P.~J.}\ \bibnamefont {R}},\ }\href
  {https://journals.aps.org/prl/abstract/10.1103/PhysRevLett.120.097701}
  {\bibfield  {journal} {\bibinfo  {journal} {Physical Review Letters}\
  }\textbf {\bibinfo {volume} {120}},\ \bibinfo {pages} {097701} (\bibinfo
  {year} {2018})}\BibitemShut {NoStop}%
\bibitem [{\citenamefont {Cristofolini}\ \emph
  {et~al.}(2012{\natexlab{a}})\citenamefont {Cristofolini}, \citenamefont
  {Christmann}, \citenamefont {Tsintzos}, \citenamefont {Deligeorgis},
  \citenamefont {Konstantinidis}, \citenamefont {Hatzopoulos}, \citenamefont
  {Savvidis},\ and\ \citenamefont {Baumberg}}]{double2d-1}%
  \BibitemOpen
  \bibfield  {author} {\bibinfo {author} {\bibfnamefont {P.}~\bibnamefont
  {Cristofolini}}, \bibinfo {author} {\bibfnamefont {G.}~\bibnamefont
  {Christmann}}, \bibinfo {author} {\bibfnamefont {S.~I.}\ \bibnamefont
  {Tsintzos}}, \bibinfo {author} {\bibfnamefont {G.}~\bibnamefont
  {Deligeorgis}}, \bibinfo {author} {\bibfnamefont {G.}~\bibnamefont
  {Konstantinidis}}, \bibinfo {author} {\bibfnamefont {Z.}~\bibnamefont
  {Hatzopoulos}}, \bibinfo {author} {\bibfnamefont {P.~G.}\ \bibnamefont
  {Savvidis}}, \ and\ \bibinfo {author} {\bibfnamefont {J.~J.}\ \bibnamefont
  {Baumberg}},\ }\href {http://science.sciencemag.org/content/336/6082/704}
  {\bibfield  {journal} {\bibinfo  {journal} {Science}\ }\textbf {\bibinfo
  {volume} {336}},\ \bibinfo {pages} {704} (\bibinfo {year}
  {2012}{\natexlab{a}})}\BibitemShut {NoStop}%
\bibitem [{\citenamefont {Li}\ \emph {et~al.}(2014)\citenamefont {Li},
  \citenamefont {Duan},\ and\ \citenamefont {Zhang}}]{double2d-1b}%
  \BibitemOpen
  \bibfield  {author} {\bibinfo {author} {\bibfnamefont {J.~Y.}\ \bibnamefont
  {Li}}, \bibinfo {author} {\bibfnamefont {S.~Q.}\ \bibnamefont {Duan}}, \ and\
  \bibinfo {author} {\bibfnamefont {W.}~\bibnamefont {Zhang}},\ }\href
  {http://iopscience.iop.org/article/10.1209/0295-5075/108/67010} {\bibfield
  {journal} {\bibinfo  {journal} {Europhysics Letters}\ }\textbf {\bibinfo
  {volume} {108}},\ \bibinfo {pages} {67010} (\bibinfo {year}
  {2014})}\BibitemShut {NoStop}%
\bibitem [{\citenamefont {Byrnes}\ \emph
  {et~al.}(2014{\natexlab{a}})\citenamefont {Byrnes}, \citenamefont {Kolmakov},
  \citenamefont {Kezerashvili},\ and\ \citenamefont {Yamamoto}}]{double2d-2}%
  \BibitemOpen
  \bibfield  {author} {\bibinfo {author} {\bibfnamefont {T.}~\bibnamefont
  {Byrnes}}, \bibinfo {author} {\bibfnamefont {G.~V.}\ \bibnamefont
  {Kolmakov}}, \bibinfo {author} {\bibfnamefont {R.~Y.}\ \bibnamefont
  {Kezerashvili}}, \ and\ \bibinfo {author} {\bibfnamefont {Y.}~\bibnamefont
  {Yamamoto}},\ }\href
  {https://journals.aps.org/prb/abstract/10.1103/PhysRevB.90.125314} {\bibfield
   {journal} {\bibinfo  {journal} {Physical Review B}\ }\textbf {\bibinfo
  {volume} {90}},\ \bibinfo {pages} {125314} (\bibinfo {year}
  {2014}{\natexlab{a}})}\BibitemShut {NoStop}%
\bibitem [{\citenamefont {Sivalertporn}\ and\ \citenamefont
  {Muljarov}(2015)}]{double2d-3}%
  \BibitemOpen
  \bibfield  {author} {\bibinfo {author} {\bibfnamefont {K.}~\bibnamefont
  {Sivalertporn}}\ and\ \bibinfo {author} {\bibfnamefont {E.~A.}\ \bibnamefont
  {Muljarov}},\ }\href
  {https://journals.aps.org/prl/abstract/10.1103/PhysRevLett.115.077401}
  {\bibfield  {journal} {\bibinfo  {journal} {Physical Review X}\ }\textbf
  {\bibinfo {volume} {115}},\ \bibinfo {pages} {077401} (\bibinfo {year}
  {2015})}\BibitemShut {NoStop}%
\bibitem [{\citenamefont {Rosenberg}\ \emph {et~al.}(2016)\citenamefont
  {Rosenberg}, \citenamefont {Mazuz-Harpaz}, \citenamefont {Rapaport},
  \citenamefont {West},\ and\ \citenamefont {Pfeiffer}}]{double2d-3b}%
  \BibitemOpen
  \bibfield  {author} {\bibinfo {author} {\bibfnamefont {I.}~\bibnamefont
  {Rosenberg}}, \bibinfo {author} {\bibfnamefont {Y.}~\bibnamefont
  {Mazuz-Harpaz}}, \bibinfo {author} {\bibfnamefont {R.}~\bibnamefont
  {Rapaport}}, \bibinfo {author} {\bibfnamefont {K.}~\bibnamefont {West}}, \
  and\ \bibinfo {author} {\bibfnamefont {L.}~\bibnamefont {Pfeiffer}},\ }\href
  {https://journals.aps.org/prb/abstract/10.1103/PhysRevB.93.195151} {\bibfield
   {journal} {\bibinfo  {journal} {Physical Review B}\ }\textbf {\bibinfo
  {volume} {93}},\ \bibinfo {pages} {195151} (\bibinfo {year}
  {2016})}\BibitemShut {NoStop}%
\bibitem [{\citenamefont {Togan}\ \emph {et~al.}(2018)\citenamefont {Togan},
  \citenamefont {Lim}, \citenamefont {Faelt}, \citenamefont {Wegscheider},\
  and\ \citenamefont {Imamoglu}}]{double2d-5}%
  \BibitemOpen
  \bibfield  {author} {\bibinfo {author} {\bibfnamefont {E.}~\bibnamefont
  {Togan}}, \bibinfo {author} {\bibfnamefont {H.~T.}\ \bibnamefont {Lim}},
  \bibinfo {author} {\bibfnamefont {S.}~\bibnamefont {Faelt}}, \bibinfo
  {author} {\bibfnamefont {W.}~\bibnamefont {Wegscheider}}, \ and\ \bibinfo
  {author} {\bibfnamefont {A.}~\bibnamefont {Imamoglu}},\ }\href
  {https://arxiv.org/abs/1804.04975} {\bibfield  {journal} {\bibinfo  {journal}
  {arxiv preprint}\ }\textbf {\bibinfo {volume} {arXiv:1804.04975}} (\bibinfo
  {year} {2018})}\BibitemShut {NoStop}%
\bibitem [{\citenamefont {Wilkes}\ and\ \citenamefont
  {Muljarov}(2017)}]{double2d-4}%
  \BibitemOpen
  \bibfield  {author} {\bibinfo {author} {\bibfnamefont {J.}~\bibnamefont
  {Wilkes}}\ and\ \bibinfo {author} {\bibfnamefont {E.~A.}\ \bibnamefont
  {Muljarov}},\ }\href
  {https://www.sciencedirect.com/science/article/pii/S0749603616314872}
  {\bibfield  {journal} {\bibinfo  {journal} {Superlattices and
  Microstructures}\ }\textbf {\bibinfo {volume} {108}},\ \bibinfo {pages} {32}
  (\bibinfo {year} {2017})}\BibitemShut {NoStop}%
\bibitem [{\citenamefont {Khadzhi}\ \emph {et~al.}(2018)\citenamefont
  {Khadzhi}, \citenamefont {Vasilieva},\ and\ \citenamefont
  {Belousov}}]{double2d-4b}%
  \BibitemOpen
  \bibfield  {author} {\bibinfo {author} {\bibfnamefont {P.~I.}\ \bibnamefont
  {Khadzhi}}, \bibinfo {author} {\bibfnamefont {O.~F.}\ \bibnamefont
  {Vasilieva}}, \ and\ \bibinfo {author} {\bibfnamefont {I.~V.}\ \bibnamefont
  {Belousov}},\ }\href
  {https://link.springer.com/article/10.1134/S1063776118020127} {\bibfield
  {journal} {\bibinfo  {journal} {Journal of Experimental and Theoretical
  Physics}\ }\textbf {\bibinfo {volume} {126}},\ \bibinfo {pages} {147}
  (\bibinfo {year} {2018})}\BibitemShut {NoStop}%
\bibitem [{\citenamefont {Rojas-Arias}\ \emph {et~al.}(2016)\citenamefont
  {Rojas-Arias}, \citenamefont {Rodr\'iguez},\ and\ \citenamefont
  {Vinck-Posada}}]{rojas-unal}%
  \BibitemOpen
  \bibfield  {author} {\bibinfo {author} {\bibfnamefont {J.~S.}\ \bibnamefont
  {Rojas-Arias}}, \bibinfo {author} {\bibfnamefont {B.~A.}\ \bibnamefont
  {Rodr\'iguez}}, \ and\ \bibinfo {author} {\bibfnamefont {H.}~\bibnamefont
  {Vinck-Posada}},\ }\href
  {http://iopscience.iop.org/article/10.1088/0953-8984/28/50/505302} {\bibfield
   {journal} {\bibinfo  {journal} {Journal of Physics: Condensed Matter}\
  }\textbf {\bibinfo {volume} {28}},\ \bibinfo {pages} {505302} (\bibinfo
  {year} {2016})}\BibitemShut {NoStop}%
\bibitem [{\citenamefont {Dory}\ \emph {et~al.}(2016)\citenamefont {Dory},
  \citenamefont {Fischer}, \citenamefont {M\"uller}, \citenamefont
  {Lagoudakis}, \citenamefont {Sarmiento}, \citenamefont {Rundquist},
  \citenamefont {Zhang}, \citenamefont {Kelaita},\ and\ \citenamefont
  {Vučković}}]{QDpolariton3}%
  \BibitemOpen
  \bibfield  {author} {\bibinfo {author} {\bibfnamefont {C.}~\bibnamefont
  {Dory}}, \bibinfo {author} {\bibfnamefont {K.~A.}\ \bibnamefont {Fischer}},
  \bibinfo {author} {\bibfnamefont {K.}~\bibnamefont {M\"uller}}, \bibinfo
  {author} {\bibfnamefont {K.~G.}\ \bibnamefont {Lagoudakis}}, \bibinfo
  {author} {\bibfnamefont {T.}~\bibnamefont {Sarmiento}}, \bibinfo {author}
  {\bibfnamefont {A.}~\bibnamefont {Rundquist}}, \bibinfo {author}
  {\bibfnamefont {J.~Y.}\ \bibnamefont {Zhang}}, \bibinfo {author}
  {\bibfnamefont {Y.}~\bibnamefont {Kelaita}}, \ and\ \bibinfo {author}
  {\bibfnamefont {J.}~\bibnamefont {Vučković}},\ }\href
  {https://www.nature.com/articles/srep25172} {\bibfield  {journal} {\bibinfo
  {journal} {Scientific Reports}\ }\textbf {\bibinfo {volume} {6}},\ \bibinfo
  {pages} {25172} (\bibinfo {year} {2016})}\BibitemShut {NoStop}%
\bibitem [{\citenamefont {Jia}\ \emph {et~al.}(2018)\citenamefont {Jia},
  \citenamefont {Schine}, \citenamefont {Georgakopoulos}, \citenamefont {Ryou},
  \citenamefont {Clark}, \citenamefont {Sommer},\ and\ \citenamefont
  {Simon}}]{QDpolariton4}%
  \BibitemOpen
  \bibfield  {author} {\bibinfo {author} {\bibfnamefont {N.~Y.}\ \bibnamefont
  {Jia}}, \bibinfo {author} {\bibfnamefont {N.}~\bibnamefont {Schine}},
  \bibinfo {author} {\bibfnamefont {A.}~\bibnamefont {Georgakopoulos}},
  \bibinfo {author} {\bibfnamefont {A.}~\bibnamefont {Ryou}}, \bibinfo {author}
  {\bibfnamefont {L.~W.}\ \bibnamefont {Clark}}, \bibinfo {author}
  {\bibfnamefont {A.}~\bibnamefont {Sommer}}, \ and\ \bibinfo {author}
  {\bibfnamefont {J.}~\bibnamefont {Simon}},\ }\href
  {https://www.nature.com/articles/s41567-018-0071-6} {\bibfield  {journal}
  {\bibinfo  {journal} {Nature Physics}\ }\textbf {\bibinfo {volume} {14}},\
  \bibinfo {pages} {550} (\bibinfo {year} {2018})}\BibitemShut {NoStop}%
\bibitem [{\citenamefont {Cristofolini}\ \emph
  {et~al.}(2012{\natexlab{b}})\citenamefont {Cristofolini}, \citenamefont
  {Christmann}, \citenamefont {Tsintzos}, \citenamefont {Deligeorgis},
  \citenamefont {Konstantinidis}, \citenamefont {Hatzopoulos},\ and\
  \citenamefont {Baumberg}}]{dipolaritons}%
  \BibitemOpen
  \bibfield  {author} {\bibinfo {author} {\bibfnamefont {P.}~\bibnamefont
  {Cristofolini}}, \bibinfo {author} {\bibfnamefont {G.}~\bibnamefont
  {Christmann}}, \bibinfo {author} {\bibfnamefont {S.~I.}\ \bibnamefont
  {Tsintzos}}, \bibinfo {author} {\bibfnamefont {G.}~\bibnamefont
  {Deligeorgis}}, \bibinfo {author} {\bibfnamefont {G.}~\bibnamefont
  {Konstantinidis}}, \bibinfo {author} {\bibfnamefont {P.~G.}\ \bibnamefont
  {Hatzopoulos}, \bibfnamefont {Z~Savvidis}}, \ and\ \bibinfo {author}
  {\bibfnamefont {J.~J.}\ \bibnamefont {Baumberg}},\ }\href {\doibase
  10.1126/science.1219010} {\bibfield  {journal} {\bibinfo  {journal}
  {Science}\ }\textbf {\bibinfo {volume} {336}},\ \bibinfo {pages} {704}
  (\bibinfo {year} {2012}{\natexlab{b}})}\BibitemShut {NoStop}%
\bibitem [{\citenamefont {Cotrino-Lemus}\ and\ \citenamefont
  {Ram\'irez}(2017)}]{cotrino}%
  \BibitemOpen
  \bibfield  {author} {\bibinfo {author} {\bibfnamefont {J.}~\bibnamefont
  {Cotrino-Lemus}}\ and\ \bibinfo {author} {\bibfnamefont {H.~Y.}\ \bibnamefont
  {Ram\'irez}},\ }\href {\doibase 10.1088/1742-6596/864/1/012081} {\bibfield
  {journal} {\bibinfo  {journal} {Journal of Physics: Conference Series}\
  }\textbf {\bibinfo {volume} {864}},\ \bibinfo {pages} {012081} (\bibinfo
  {year} {2017})}\BibitemShut {NoStop}%
\bibitem [{\citenamefont {Byrnes}\ \emph
  {et~al.}(2014{\natexlab{b}})\citenamefont {Byrnes}, \citenamefont {Kim},\
  and\ \citenamefont {Yamamoto}}]{bose}%
  \BibitemOpen
  \bibfield  {author} {\bibinfo {author} {\bibfnamefont {T.}~\bibnamefont
  {Byrnes}}, \bibinfo {author} {\bibfnamefont {N.~Y.}\ \bibnamefont {Kim}}, \
  and\ \bibinfo {author} {\bibfnamefont {Y.}~\bibnamefont {Yamamoto}},\ }\href
  {\doibase https://doi.org/10.1038/nphys3143} {\bibfield  {journal} {\bibinfo
  {journal} {Nature Physics}\ }\textbf {\bibinfo {volume} {10}},\ \bibinfo
  {pages} {803} (\bibinfo {year} {2014}{\natexlab{b}})}\BibitemShut {NoStop}%
\end{thebibliography}%

\end{document}